\documentclass[twocolumn]{aastex63}

\usepackage[frozencache, cachedir=.]{minted}
\usepackage{graphicx}
\usepackage{amsmath}
\defcitealias{venot2020}{Venot2020}

\received{June 1, 2019}
\revised{January 10, 2019}
\accepted{\today}
\submitjournal{AJ}

\shorttitle{\textsc{FRECKLL}}
\shortauthors{Al-Refaie, Venot, Changeat and Edwards (2022)}
\graphicspath{{./}{figures/}}
\begin{document}

\title{\textsc{FRECKLL}: Full and Reduced Exoplanet Chemical Kinetics
               distiLLed}

\correspondingauthor{Ahmed Faris Al-Refaie}
\email{ahmed.al-refaie.12@ucl.ac.uk}

\author[0000-0003-2241-5330]{Ahmed Faris Al-Refaie}
\affil{Department of Physics and Astronomy, University College London, Gower Street, London, WC1E 6BT, UK}

\author[0000-0003-2854-765X]{Olivia Venot}
	\affil{Universit\'{e} de Paris Cit\'{e} and Univ Paris Est Creteil,
		CNRS, LISA,
		F-75013 Paris, France}

\author[0000-0001-6516-4493]{Quentin Changeat}
\affil{European Space Agency (ESA), ESA Office, 
        Space Telescope Science Institute (STScI), Baltimore MD 21218, USA}
\affil{Department of Physics and Astronomy, University College London, Gower Street, London, WC1E 6BT, UK}

\author[0000-0002-5494-3237]{Billy Edwards}\thanks{Paris Region Fellow}
\affil{AIM, CEA, CNRS, Universit\'e Paris-Saclay, Universit\'e de Paris, F-91191 Gif-sur-Yvette, France}
\affil{Department of Physics and Astronomy, University College London, Gower Street, London, WC1E 6BT, UK}


\begin{abstract}

We introduce a new Python 1D chemical kinetic code \textsc{FRECKLL} (Full and Reduced Exoplanet Chemical Kinetics distiLLed) to evolve large chemical networks efficiently. \textsc{FRECKLL} employs `distillation' in computing the reaction rates, which minimizes the error bounds to the minimum allowed by double precision values ($\epsilon \leq 10^{-15}$). Compared to summation of rates with traditional algorithms like pairwise summation, distillation provides a tenfold reduction in solver time for both full and reduced networks. Both the full and reduced \citetalias{venot2020} networks are packaged in \textsc{FRECKLL} as well as a \textsc{TauREx} 3.1 plugin for usage in forward modelling and retrievals of exoplanet atmospheres. We present \textsc{TauREx} retrievals performed on a simulated HD189733 JWST spectra using the full and reduced \citetalias{venot2020} chemical networks and demonstrate the viability of total disequilibrium chemistry retrievals and the ability for JWST to detect disequilibrium processes.

\end{abstract}


\section{Introduction}

In the last decade, observations from space using mainly the Hubble Space Telescope (HST) and the Spitzer Space Telescope (Spitzer) and from the ground, have allowed to characterise the atmospheric properties of a handful of planets from their transit \citep{tinetti, Kreidbger_2014, Sing_2016, Sedaghati_2017, Wakeford_2017, Tsiaras_2018, Fisher_2018, Anisman_2020, Edwards_2021, Gressier_2022, Wong_2022, Saba_2022, Edwards_2022}, eclipse \citep{Swain_2008, Crouzet_2014, Line_2014, Haynes_2015, Line_2016, Edwards_2020, Changeat_2021_K9, Changeat_2022_five, Fu_2022} or phase-curve observations \citep{Stevenson_2017_W43, Arcangeli_2019, Changeat_2021_pc2, Changeat_2022, Evans_2022, Chubb_2022}. Due to the low resolution and narrow wavelength coverage of older generation space-based instrumentation, however, degeneracies can often lead to multiple interpretations of exoplanet spectra, depending on model and prior assumptions \citep[e.g.][]{Changeat_2020_K11}. To explore the information contained in these spectra, exoplanet teams have developed sophisticated methods to invert the information content in the spectra of exoplanets. These methods, often called spectral retrieval techniques \citep{Irwin_2008, Madhu_2009, Benneke_2012, Line_2013, Waldmann_2015, Min_2020, Moliere_2019, Al-Refaie_2021} require the evaluation of thousands to millions of forward models, therefore requiring significant computing resources. Often, the computing requirements imply that simplified atmospheric models have to be employed, for instance, by assuming 1-dimensional geometries and other idealized assumptions on the thermal structure, the chemistry and the cloud properties. Since the information extracted from current spectra is low, assumptions are commonly used throughout the literature. These assumptions include isothermal thermal structure, constant chemical profiles or equilibrium chemistry, and fully opaque cloud opacities.

In retrieval codes, the chemistry is often recovered using profiles constant with altitude; a single free parameter representing each molecule (e.g., see references above). While not representative of an entire atmosphere, current observations mostly probe small pressure regions where chemical variations remain small. An alternative assumption is thermochemical equilibrium \citep{white_1958, Eriksson1971}, which requires computing the chemistry state by minimizing the Gibbs free energy of the system. Such an assumption has gained popularity due to the reduced degrees of freedom. Furthermore, it often only requires two free parameters for metallicity and the C/O ratio chosen for their natural links to planetary formation and evolution processes. Equilibrium chemistry, however, is a strong assumption with little justification. Given with \citet{Tsai2023} and \citet{Dyrek2024}, in which SO$_2$, produced by photolyses, has been detected, equilibrium does not represent the underlying chemical processes being detected in exoplanet atmospheres. Furthermore, simulations employing kinetics methods and thus taking into account disequilibrium processes such as mixing and photochemistry have proven that chemical equilibrium is inadequate in many scenarios (e.g. \citealt{Moses_2011, Moses_2013, Moses_2016, Venot_2012, Venot_2014, venot2020,Venot_2020_W43, Molaverdikhani_2019, Tsai_2021, Morley_2017, molliere_2020, kawashima_2021}). With future telescopes, accurate representation of the chemical processes will be essential to ensure unbiased interpretation of the observations as discussed in the first analyses of JWST data \citep{ERS_2022, Tsai2023, Dyrek2024}.

In this paper we present the first implementation of a full chemical kinetic scheme into an atmospheric retrieval framework. We use the flexibility of the plugin system in \textsc{TauREx} 3.1 to integrate this new scheme and explore the use of chemical kinetic models in atmospheric retrievals. In particular, we focus our study on quantifying the impact of the equilibrium chemistry assumption in interpreting atmospheres exhibiting disequilibrium processes. Section 2 presents our implementation of the chemical kinetic code and the steps carried out in this work. In Section 3, we present the results of our simulations. Finally, Section 4 discusses our findings and provides the main conclusions of our exploration. 

\section{Kinetic Model}

\subsection{Description of chemical kinetic model}

As opposed to thermochemical equilibrium models, which predict the chemical state of a planet's atmosphere by minimising the Gibbs free energy of the system, chemical kinetic models necessitate integrating the system of differential equations representing each considered reaction until a steady state is reached. The continuity equation (Equation \ref{eq:diseq}) describes the temporal evolution of the abundance of each species $i$, considering a one-dimensional plane-parallel atmosphere. 
\begin{equation} \label{eq:diseq}
\frac{\partial n_i}{\partial t} = P_i -L_i - \frac{\partial \phi_i}{\partial z}
\end{equation}
where $n_i$, $P_i$ and $L_i$ are the number density (cm$^{-3}$), production rate and loss rate of species $i$ (cm$^{-3}$.s$^{-1}$), $z$ is the vertical coordinate of the atmosphere, and $\phi_i$ is the vertical flux for species $i$ which has the form of a diffusion equation given in Equation \ref{eq:diffu}.
\begin{equation} \label{eq:diffu}
\phi_i = -n_i D_i \left( \frac{1}{n_i}\frac{\partial n_i}{\partial z} + \frac{1}{H_i} + \frac{1}{T}\frac{\partial T}{\partial z}\right) - n_i K_{zz}\left(\frac{1}{y_i}\frac{\partial y_i}{\partial z}\right) 
\end{equation}
Here, $D_i$ is the molecular diffusion coefficient (cm$^2$ s$^{-1}$), $H_i$ is the scale height (km) and $y_i$ the mixing ratio for species $i$, T is the temperature (K) and $K_{zz}$ is the eddy diffusion coefficient (cm$^2$ s$^{-1}$). While, rigorously, thermal diffusion should be included in this equation \citep{drummond_2016}, it has been found that it was negligible compared to the other terms of Equation \ref{eq:diffu} \citep{venot_phd}.  We thus don't include it.

At $t=0$ s, an initial abundance is set. \textsc{FRECKLL} can accept any initial atmospheric abundance, either user-supplied or from an external code. As the default, the system is initialised with the abundance of each species assumed to be at thermochemical equilibrium. This initial state is computed using the ACE code \citep{Agundez2012} with supplied or user-defined NASA polynomial thermochemical coefficients and, subsequently, Equation \ref{eq:diseq} is evolved using a stiff ODE solver such as VODE \citep{vode} or DLSODES from the ODEPACK \citep{odepack} package until steady state is achieved or a user-defined condition is reached. Metallicity, elemental ratios (e.g. C/O and N/O ratios) can be set to determine the initial abundance produced by ACE. In addition to those parameters, the model also allows for the definition of the eddy diffusion parameter $K_{zz}$, given as a constant value or layer-by-layer. For our test cases, we employ the full \citet{venot2020} network. This network is based on the \citet{Venot_2012} network with updates to its methanol chemistry and includes 108 species, 1906 reactions and 55 photodissociations (see Table \ref{tab:photo-reactions}). The network, including all the photolyses data can be found on the website of the ANR EXACT\footnote{\url{https://www.anr-exact.cnrs.fr/}}. In addition, \citet{venot2020} provides a reduced network consisting of 44 species and 582 reactions and omits photolysis reactions.

For a 130 layer atmosphere, \textsc{FRECKLL} takes roughly 4--5 minutes to reach a steady-state, using the full chemical network and roughly 30 seconds on the reduced, significantly speeding up convergence.
% Included is another chemical model that exploits the quench approximation\citep{prinnbarshay1977}. Rather than attempting to fully evolve Equation \ref{eq:diseq} we, instead, determine a 'quench point' in the atmosphere where chemical and dynamical timescales are equal. Below this point, chemical equilibrium is assumed, above this point abundances are frozen at the equilibrium value of the quench point. The chemical timescale is computed as such:

% \begin{equation} \label{eq:chemtime}
% t_{chem} = n_i(\frac{\partial n_i}{\partial t})^{-1}
% \end{equation}
% and the dynamical timescale:
% \begin{equation} \label{eq:chemtime}
% t_{dyn} = \frac{L^{2}}{K_{zz}}
% \end{equation}
% where $L^{2}$ is the characteristic scaleheight, taken as the pressure scaleheight $L=\alpha H$ where $\alpha$ is a scaling factor between 0.1--1.0 \citep{smith98,vissmoses11}. By default $\alpha=1$ but can be included in retrievals as a fittable parameter.

\subsection{The importance of Numerical stability}

Floating points, being approximations of real numbers, inherently carry errors with each operation. Numerical stability involves understanding, managing, and minimizing these errors to ensure the accuracy and reliability of computations. This problem is not exclusive to Python or the particular packages highlighted in this paper. These challenges permeate various computational platforms, even extending to compiled languages like FORTRAN. Chemical kinetics, by nature, presents inherent stiffness in its equations, stemming from the wide-ranging magnitudes of chemical timescales and abundances. Integration requires stiff ODE algorithms such as Backwards differentiation formula, Rosenbrock and backwards Euler methods which can vary the time steps over large orders of magnitude. Additionally, an overlooked aspect involves the computation of sums. With double precision we generally expect the upper bound of relative errors from rounding to be $\epsilon_{m}\approx10^{-16}$. Assuming a function implemented with algorithm $f(x)$ and the true function $\tilde{f}(x)$ the relative error for an algorithm $\epsilon$ is computed as:
\begin{equation}
    \epsilon = \frac{|f(x) - \tilde{f}(x)|}{|\tilde{f}(x)|}
\end{equation}
Here, the true function is $\tilde{f}(x) = \sum^n x_i$ where summation is performed at infinite precision. We must also consider the condition number $C$:
\begin{equation}
    C = \frac{\sum^n{|x_i|}}{|\sum^n{x_i}|}
\end{equation}
which represents the intrinsic sensitivity of summation. For naive summation such as the inbuilt python \texttt{sum} function, the error is bounded as:
\begin{equation}
    \epsilon \leq n\epsilon_{m}C
\end{equation}
where $n$ is the number of elements. The error for pairwise summation used by the \texttt{numpy.sum} function is bounded by:
\begin{equation}
    \epsilon \leq \frac{\epsilon_{m}log_2 n}{1-\epsilon_{m}log_2 n}C
\end{equation}

Generally, well-conditioned problems are those where $C\approx1$, one such case is where all values are non-negative (i.e $x_i > 0$). For 10,000 elements, the error from naive summation is $\epsilon \leq 10^{-12}$ and for pairwise we expect an error $\epsilon \leq 10^{-15}$.

The problem comes when dealing with extensive magnitudes and a mixture of negative and non-negative values.
To illustrate, let us take an array of values $x = [10^{16}, 10^{30}, 1, 5, 10, 10^{4}, -10^{30}, -10^{16}]$, where $\sum x = 10016$. Attempting to use the native sum we get:
\begin{minted}{python}
>>> sum( [1e+16, 1e+30, 1, 5, 10, 
          10000.0, -1e+30, -1e+16])
-7638326771712.0
\end{minted}
the error is in the order of $\epsilon \approx 10^{8}$. Pairwise summation performs a little better:
\begin{minted}{python}
>>> numpy.sum( [1e+16, 1e+30, 1, 5, 
                10, 10000.0, -1e+30, -1e+16])
0.0
\end{minted}
here $\epsilon = 1$. The problem is \textit{ill-conditioned} with a condition number of $C=10^{26}$ which is extremely large. This arises from \textit{catastrophic cancellation} where precision limits for floating points mean $dbl(10^{30} + 1) = 10^{30}$ where $dbl$ is an operation under double-precision. For kinetics calculations, this is problematic, as summing production and loss rates for a molecule can suddenly become zero, change sign or magnitude. In the Jacobian, these appear as sudden discontinuities and can cause stiff ODE methods to oscillate at certain times and continually reduce the time-step, which will drop the integration efficiency.
To avoid these problems in \textsc{FRECKLL}, we instead employ the K-fold summation method \citep{Ogita2005}. This method first performs an error-free transformation of an array:

\begin{minted}{python}
def twosum(a,b):
    x = a + b
    z = x - a
    y = (a - (x-z)) + (b-z)
    return x,y
    
def vecsum(x):
    for i in range(1, len(x)):
        a,b = twosum(x[i], x[i-1])
        x[i] = a
        x[i-1] = b
    return x
\end{minted}

The \texttt{twosum} computes the resultant floating point sum and residuals from the summation. For each element $i$, the result is stored at $i$ and the residual at $i-1$. For an array $x$ the algorithm produces a resultant array $y$ where $y = vecsum(x)$ which has the property:
\begin{equation}
    \sum^{n} y_i = \sum^{n} x_i
\end{equation}
assuming infinite precision. This is often referred to as `distillation' \citep{kahan}. For low condition numbers, elements $i=1...n-1$ will be zero and $i=n$ will contain the resultant sum (i.e $y_n = \sum^{n} x_i$). For higher condition numbers this takes the form $\sum_{n-1} y_i + y_n = \sum x_i$. Distillation has the effect of reducing the condition number and indeed, applying it to our original array $x = [10^{16}, 10^{30}, 1, 5, 10, 10^{4}, -10^{30}, -10^{16}]$, the condition number falls to $C\approx10^{9}$. As distillation preserves the original sum and reduces the condition number, we can apply it $K-1$ times until we reach our desired condition number before applying the summation; this is K-fold summation in its essence:
\begin{minted}{python}
def kfold(x, K):
    v = x
    for k in range(K-1):
        v = vecsum(v)
    
    return sum(v[:-1]) + v[-1]

>>> kfold( [1e+16, 1e+30, 1, 5, 10,
            10000.0, -1e+30, -1e+16], K=2)
10016.0
\end{minted}
Applying this algorithm for $K=2$ we indeed get the correct result. The error bounds for $K=2$ are $\epsilon_{K=2} \leq 10^{4}$. Increasing to $K=4$ gives an error bounded $\epsilon_{K=4} \leq 10^{-15}$ which is the maximum possible with double precision. K-fold summation is significantly slower than \texttt{numpy.sum}; 10,000 elements takes roughly 7-10x longer than \texttt{numpy}. However, as we will demonstrate, the increase in precision greatly benefits convergence.
We employ K-fold summation in computing the production and loss rates of molecules. To maximise computational efficiency and precision we combine the rates into a single array. For molecule $i$ and reaction $r$ we combine the production rate $P^{i}_r$ and loss rate $L^{i}_r$ into a total molecule rate $R^{i}$. If we have $p$ production reactions and $l$ loss reactions then:
\begin{equation}
\begin{split}
R^{i}_{1...p} &= P^{i}_{1...p}\\
R^{i}_{p+1...p+l} &= -L^{i}_{1...l} \\
\end{split}
\end{equation}

\begin{center}
\begin{figure}[]
    \includegraphics[width=1.0\columnwidth]{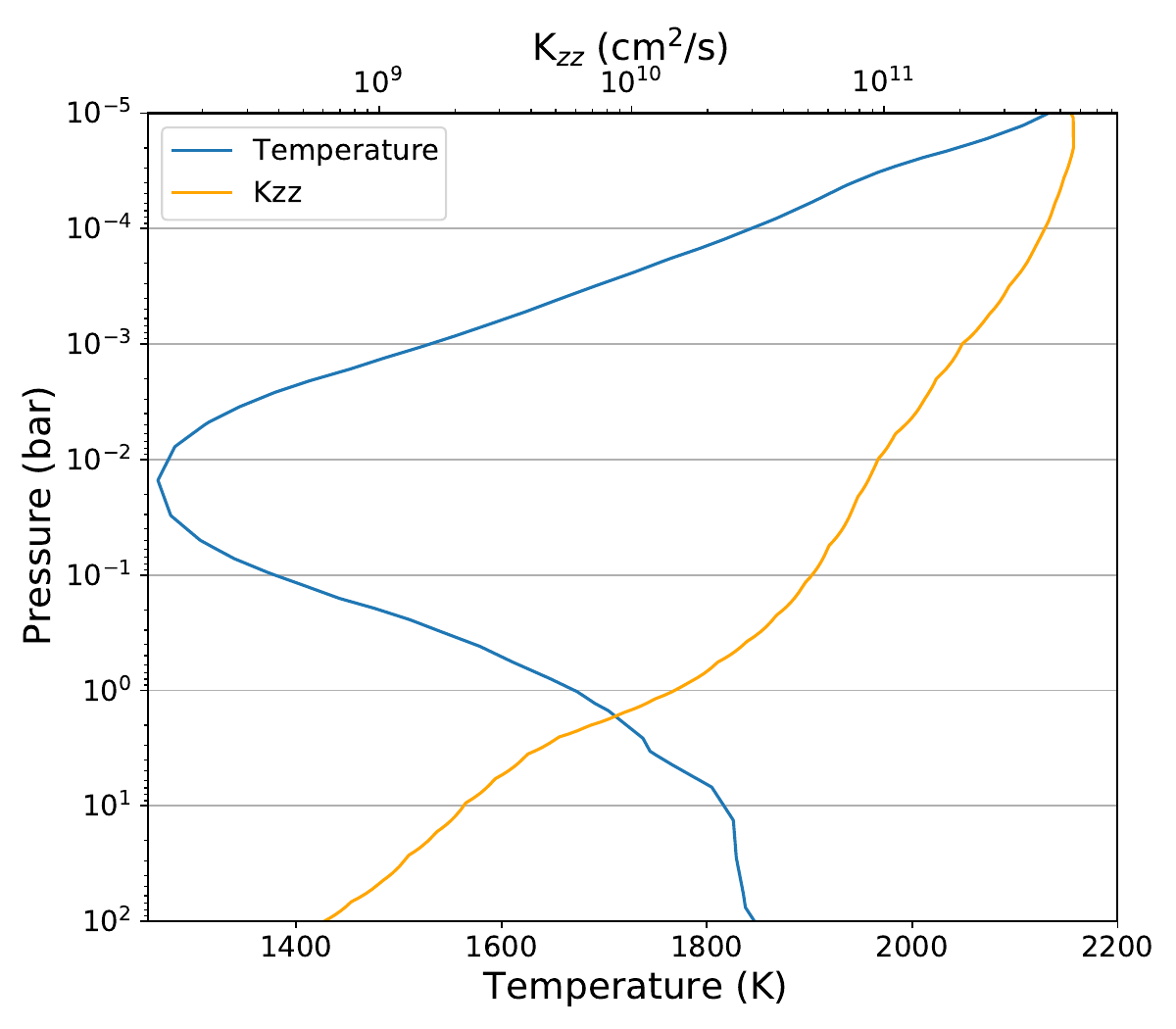}
\caption{Temperature and vertical mixing ($K_{zz}$) profiles of HD\,209458\,b taken from \cite{venot2020}.}
\label{fig:profiles}
\end{figure}
\end{center}

The total rate for molecule $i$ is given as:
\begin{equation}
    R_i = k(R^{i}_r, K=4)
\end{equation}
where $k$ is our K-fold function. We can rewrite Equation \ref{eq:diffu} as:
\begin{equation}
\frac{\partial n_i}{\partial t} = R_i - \frac{\partial \phi_i}{\partial z}
\end{equation}
Additionally we also include convergence criteria to numerically determine if steady-state has been reached. The criteria is the same as the one used in \textsc{VULCAN} \citep{vulcan}. Given timesteps $m$ and $m+1$ we compute the following:
\begin{equation}
\begin{split}
\Delta n_i &= \frac{n_{i,m+1} - n_{i,m}}{n_{i,m}} \\
\Delta t &= t_{m+1} - t_{m} \\ 
\end{split}
\end{equation}
Our criteria for steady state is therefore:
\begin{equation}
\begin{split}
\max{|\Delta n_i|} &< \delta \\
\max{|\frac{\Delta n_i}{\Delta t}|} &< \eta \\
\end{split}
\end{equation}
where $\delta$ and $\eta$ are our criterion parameters for the relative change and relative change over time, respectively. We demonstrate the benefit of K-fold summation by solving a benchmark system. We compute a HD\,209458\,b model between $10^{-5}$--$10^{2}$ bar, using the thermal and vertical mixing profiles from \cite{venot2020} displayed in Figure \ref{fig:profiles}. The model consists of 130 layers, 108 molecules, 1906 reactions and 55 photodissociations. For the actinic flux, as HD 209458 is a G0 star, we use the UV spectral irradiance of the Sun \citep{Thuillier2004} scaled to correspond to the radius and effective temperature of HD 209458. We evolve the system using the VODE solver with a relative tolerance of $10^{-3}$ and absolute tolerance at $10^{-25}$ until $t=10^{10}$ s with steady-state occurring at $t=10^{8}$ s. For this case we set $\delta=0$ and $\eta=0$, disabling the criteria. We will also assess the number of function evaluations (evaluation of Eq. \ref{eq:diseq}) and the number of times the jacobian matrix is evaluated.

\begin{center}
\begin{figure}[]
\begin{interactive}{animation}{end_result.mp4}
    \includegraphics[width=1.0\columnwidth]{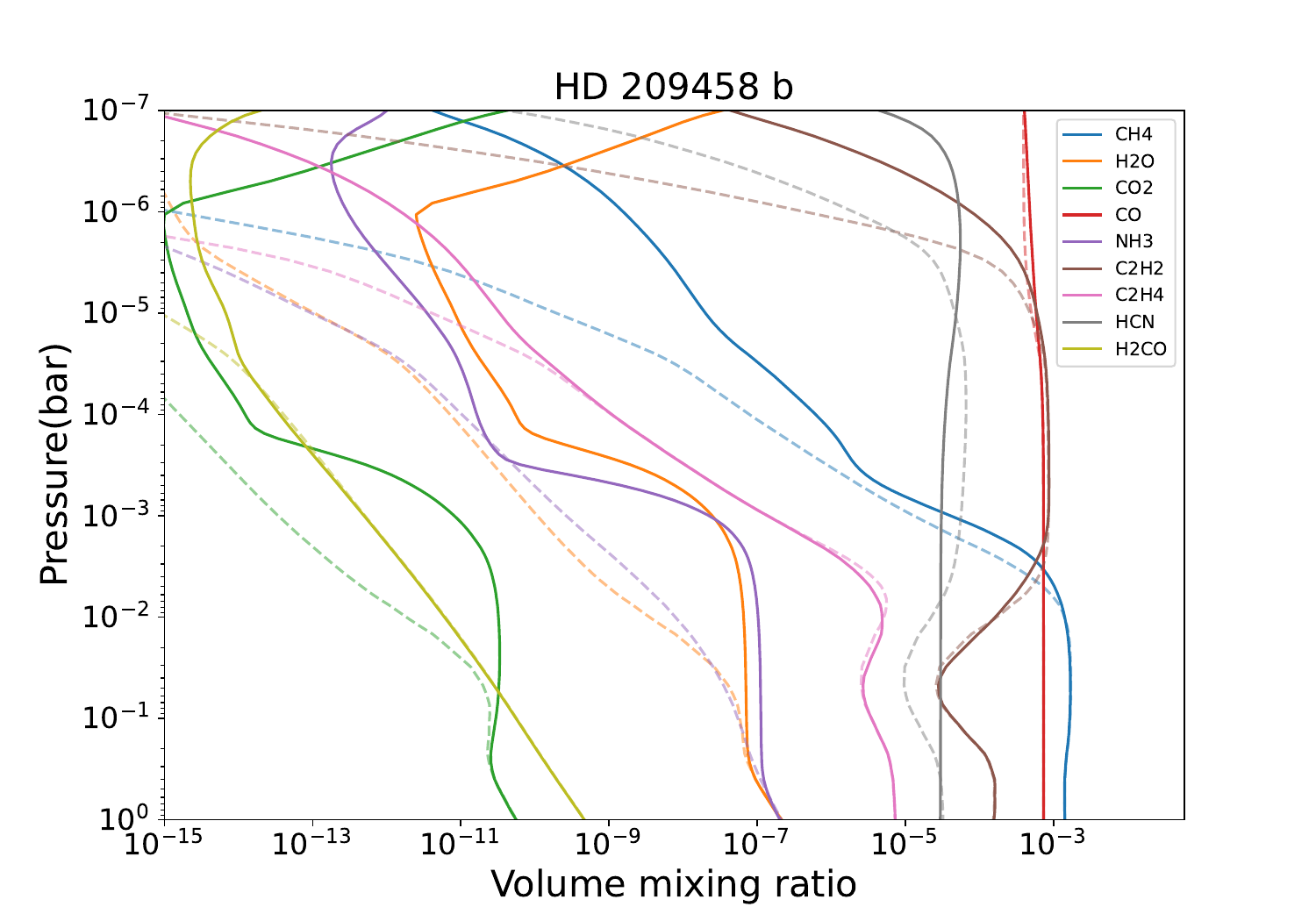}
\end{interactive}
\caption{Chemical abundances for a benchmark HD 209458 b atmosphere using temperature and vertical mixing profiles as well as the full chemical network from \cite{venot2020}. The dashed lines are the system's initial state at equilibrium abundances and the solid lines the final steady state solution at $t=10^{10}$s. An animated version of this figure describes the same plot but evolving from $t=0$s to steady state at $t=10^{10}$s.}
\label{fig:hd209-kinetic}
\end{figure}
\end{center}

Figure \ref{fig:hd209-kinetic} shows the initial abundances at equilibrium computed using ACE and the final steady state solution achieved by \textsc{FRECKLL} at $t=10^{10}$ s.
Using pairwise summation implemented by \texttt{numpy.sum} takes roughly 128 minutes to evolve until $t=10^{10}$ s requiring 467335 function evaluations and 2179 jacobian evaluations. The issue is that the solver is unable to take larger time steps, especially at $t\approx 10^{8}$ s where $\Delta t \approx 10^{4}$ s. As discontinuities appear more often, the solver is forced to use small timesteps in order to ensure smoothness in the function. This effect becomes more pronounced as the system approaches steady state as the solver has difficulty integrating $|\frac{\partial n_i}{\partial t}|$ below a certain threshold. Tracing this during integration estimates this threshold to be around $10^{-10}$.

% Worryingly, examining the evolution history at $t > 10^{8}$, numerical precision issues causes it to overshoot steady state as seen in Figure \ref{fig:evo_piecewise}. Whilst prior knowledge has allowed us to carefully chose the integration regions to avoid this and ensure steady state at an integration boundary, other systems can unintentionally encounter this behaviour.
Solving the same system using K-fold summation ($K=4$) until $t=10^{10}$ s takes 5 minutes, 2,682 function evaluations and 158 jacobian evaluations. As we stated previously, K-fold summation is significantly slower than piecewise summation but we manage to gain a 25x reduction in solver time as well as a 175x reduction in function evaluations. The improved precision means that $|\frac{\partial n_i}{\partial t}|$ can reach $10^{-15}$ and the solver is choosing larger time-steps that skip from $10^{8}$ s$-10^{10}$ s. In fact, solving further to $10^{12}$ s takes only 10 extra function evaluations.

There is always a trade-off between raw performance and precision. When dealing with stiff non-linear systems, convergence can be hampered by the underlying precision of algorithms. It is sometimes easy to forget that summation is also an algorithm and not an intrinsic feature of computation. We demonstrate that choosing a slower, more precise summation algorithm can lead to significant performance gains from faster convergence. We also like to note that the summation algorithm presented here is not exclusive to chemical kinetics and can be applied to other ill-conditioned problems.

% \section{TauREx plugin}

\section{Forward Models}

\textsc{FRECKLL} includes a plugin for \textsc{TauREx} 3.1 \citep{Al-Refaie_2021, Al-refaie_2021_t3.1} for generation of synthetic spectra and retrievals using the chemical kinetic code. We demonstrate its forward modelling capabilities by simulating HD\,189733\,b with parameters taken from the literature which are given in Table \ref{tab:hd189-test}. Note that we simulate HD\,189733\,b with a constant $K_{zz}$ of 4$\times$10$^8$ cm$^2$s$^{-1}$. This is a fairly large value for the eddy diffusion coefficient, suggesting strong vertical mixing in the atmosphere in this scenario. For simplicity, the temperature profile for those simulations is modelled using an isothermal profile, even if GCM models predict variations with altitude, as well as with longitude and latitude \citep[e.g.][]{Drummond_2020}. In the model, we include absorption using the ExoMol line-lists \citep{Tennyson_2012_genericExomol, Chubb_2021_genericExomol, Tennyson_2020_genericExomol} from the species H$_2$O \citep{polyansky_2018_h2o}, CH$_4$ \citep{Yurchenko_2014_ch4}, CO \citep{Li_2015_co}, CO$_2$ \citep{Yurchenko_2020_co2}, NH$_3$ \citep{Coles_2019_nh3}, HCN \citep{Harris_2006_hcn}, C$_2$H$_2$ \citep{Chubb_2020_c2h2}, C$_2$H$_4$ \citep{Mant_2018_c2h4} and H$_2$CO \citep{al-refaie_2015_h2co}. We also include Collision Induced Absorption from H$_2$-H$_2$ \citep{abel_h2-h2, fletcher_h2-h2} and H$_2$-He \citep{abel_h2-he} and Rayleigh Scattering for H$_2$, He, N$_2$, O$_2$, CO$_2$, CH$_4$, CO, NH$_3$ and H$_2$O given by \cite{cox_allen_rayleigh}. The atmosphere is modelled in plane-parallel geometry with 100 layers spaced between 10 bar and 10$^{-5}$ bar in log space. The actinic flux used for HD 189733 is the same as the one used in \citet{Venot_2012}, originally produced by Ignasi Ribas (priv. comm.), it comprises of spectra from X-exoplanets \citep{Forcada2011}, FUSE and HST data of $\epsilon$ Eridani and PHOENIX data \citep{Hauschildt1999} for the spectral regions 0.5--90~nm, 90--330~nm and 330+~nm respectively.
We solve the kinetics with a relative tolerance of 10$^{-3}$ and we reduce the absolute tolerance to 10$^{-20}$ for speed without harming the precision of the retrieval as molecules below this density are not spectrally visible. We set the convergence criteria to $\delta=10^{-4}$ and $\eta=10^{-4}$ and maximum integration time of $t=10^{30}$s. On a 2.3 GHz Quad-Core Intel Core i5, the single-core combined runtime (kinetics and radiative transfer) for the reduced and full networks are 23 seconds and 3.2 minutes, respectively.

\begin{center}
\begin{figure}[]
    \includegraphics[width=1.0\columnwidth]{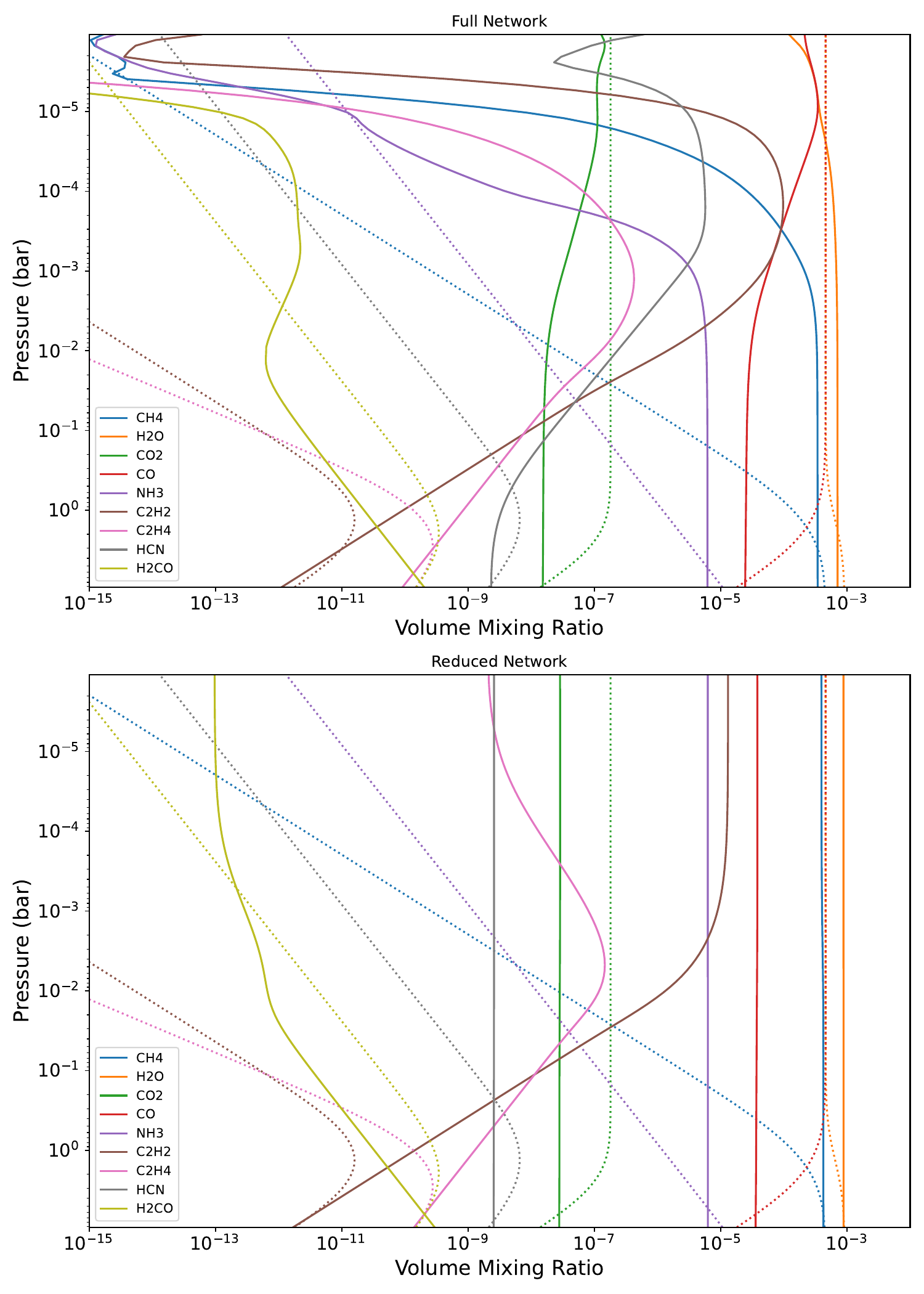}
\caption{Vertical abundance profiles for the main constituents of HD\,189733\,b computed with \textsc{FRECKLL} using the Full \citetalias{venot2020} network (top) and the Reduced \citetalias{venot2020} network (bottom). Dotted lines are equilibrium molecular profiles. Planetary and star parameters used are from Table \ref{tab:hd189-test}.}
\label{fig:forward_chem}
\end{figure}
\end{center}

\begin{center}
\begin{figure}[]
    \includegraphics[width=1.0\columnwidth]{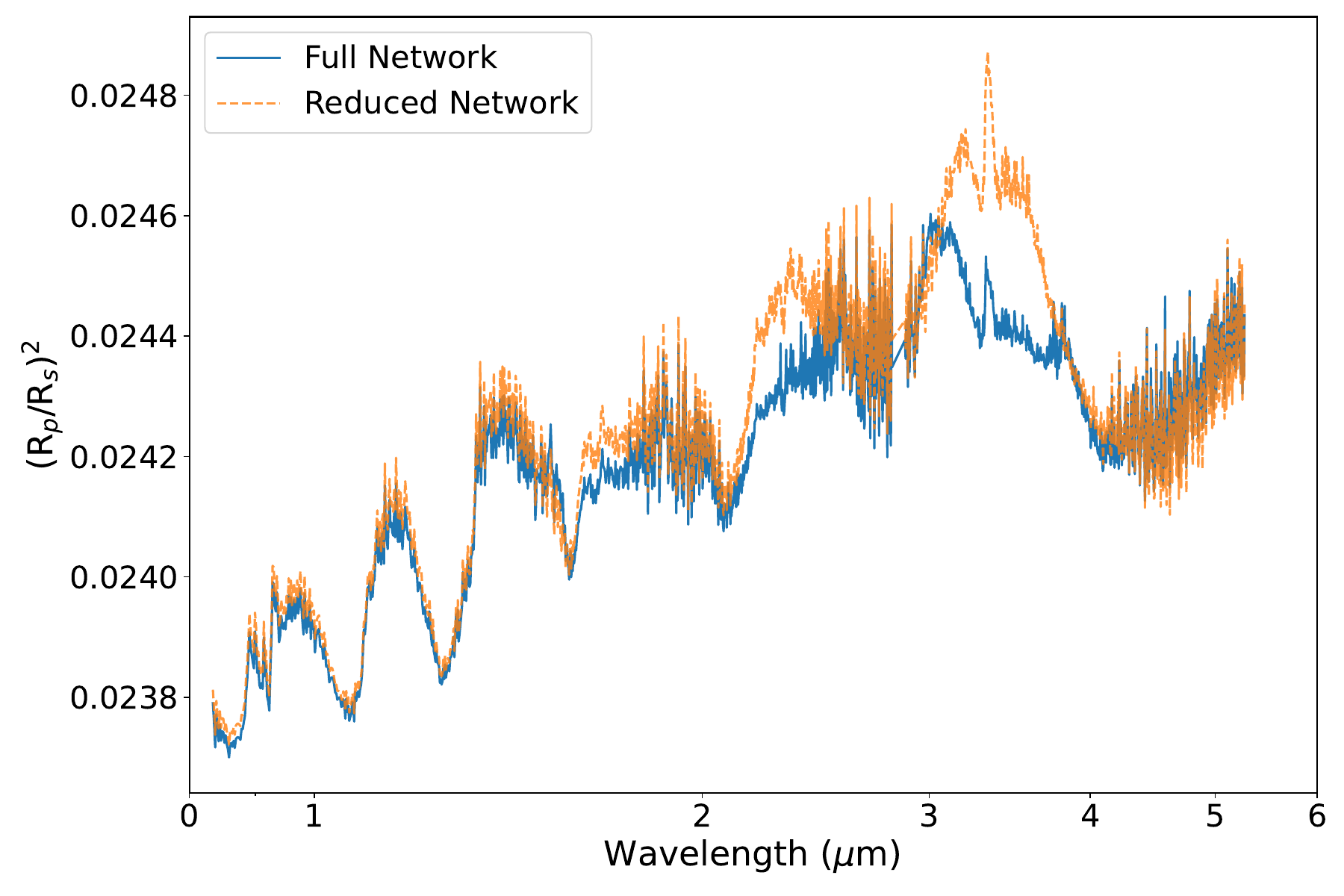}
\caption{Synthetic transmission spectra for HD 189733 b computed with \textsc{TauREx} of HD\,189733\,b using parameters from Table \ref{tab:hd189-test} and the results of the two forward models computed with FRECKLL.}
\label{fig:forward_spec}
\end{figure}
\end{center}

The chemical profiles from the reduced and full networks are presented in Figure \ref{fig:forward_chem} with corresponding transmission spectra given in Figure \ref{fig:forward_spec}.

\begin{table}

\centering
\begin{tabular}{llr}
\hline\hline
Parameter & Description & Value \\
\hline 
HD 189733 &  & \\
\hline
$R_s$ & Stellar radius & 0.76 $R_{\odot}$\\
$T_s$ & Stellar temperature & 5050.0 K \\
$K_{mag}$ & K-band magnitude & 5.541 \\
$D_s$ & Distance to the star & 27 pc$^{*}$ \\
$Z_s$ & Stellar metallicity & 0.01 $Z_{\odot}$ \\
$M_s$ & Stellar mass & 0.82 $M_{\odot}$ \\
\hline
HD 189733 b & &\\
\hline
$R_p$ & Planetary radius & 1.12 $R_{J}$ \\
$M_p$ & Planetary mass & 1.16 $M_{J}$ \\
Semi-major axis &  & 0.031 AU \\
$t_{period}$ & Orbital period & 2.219 days \\
$t_{transit}$ & Transit Duration & 1.84 hours \\
$T$ & Effective temperature & 1200 K \\
$Z$ & Planetary metallicity & $Z_{\odot}$ \\
C/O &  & 0.5 \\
$K_{zz}$ (log10) & Eddy diffusion coeff. & 4x10$^{8}$ cm$^2$/s  (8.60) \\
\hline
\end{tabular}
\caption{Planetary and parent star parameters from \citet{addison2019} used to generate the simulated HD\,189733\,b JWST transit spectra. $^{*}$ we've elected to move the star further away to prevent saturation of the JWST NIRISS instrument. K$_{zz}$ includes the log10 value in brackets.}
\label{tab:hd189-test}
\end{table}

\begin{center}
\begin{figure}[]
    \includegraphics[width=1.0\columnwidth]{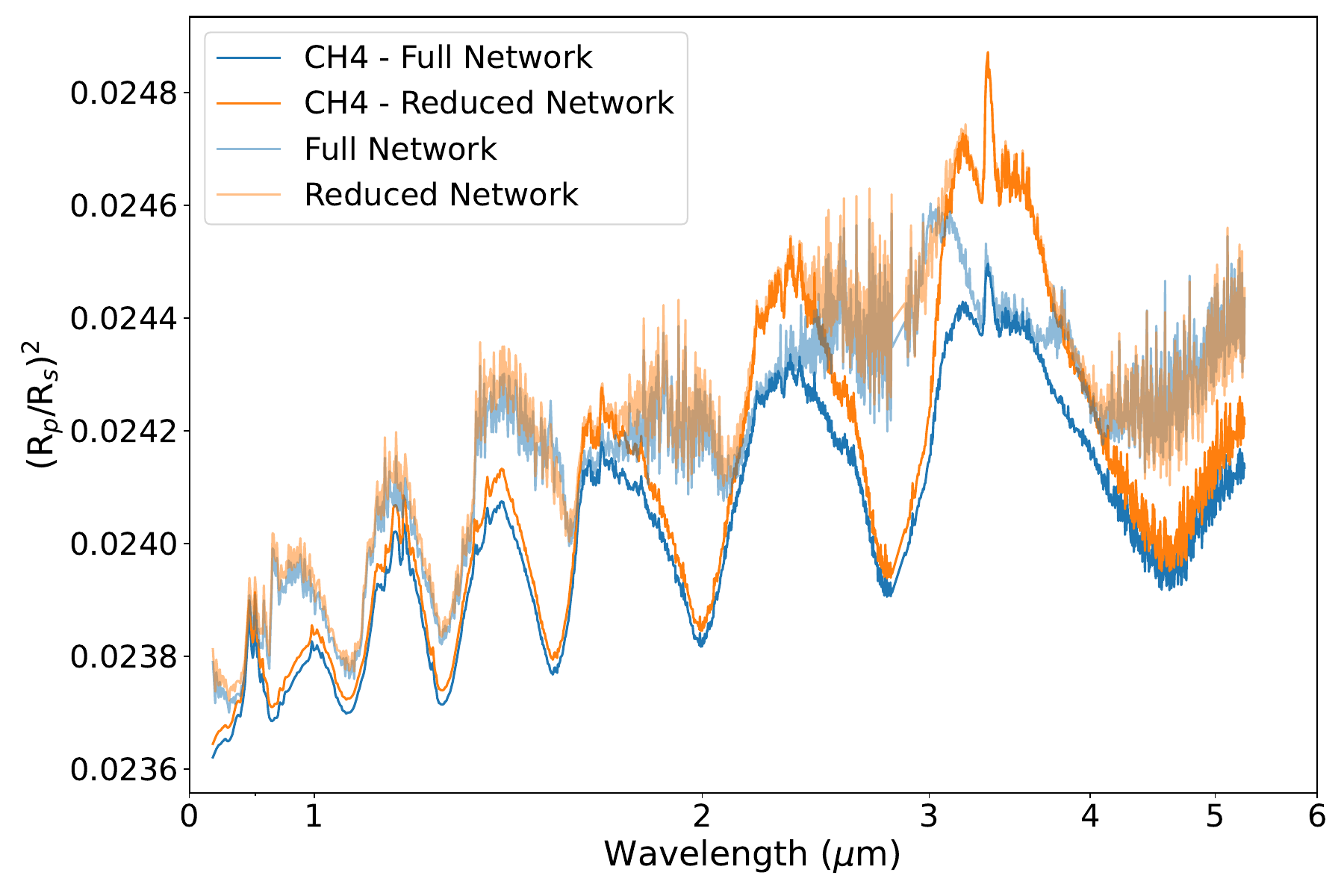}
\caption{Contribution of methane for both full (solid blue plot) and reduced (solid orange plot) networks compared to their corresponding HD\,189733\,b spectra (shaded lines) using parameters from Table \ref{tab:hd189-test}.} 
\label{fig:ch4-contrib}
\end{figure}
\end{center}

We observe large differences between the two chemical networks from the chemistry predictions. In particular, while the predictions at the bottom of the atmosphere are consistent, large differences in the predicted abundances can be seen for the top of the atmosphere (below 0.1 bar). Those differences are likely due to the inclusion of reactions for photochemistry in the full network. We note, in particular, that the abundances of CH$_4$ and NH$_3$ decrease very rapidly for pressures above 10$^{-3}$ bar, while the C$_2$H$_2$ profile is significantly affected. This translates in large differences in the observed spectrum at the wavelengths that are probing those altitudes. For instance, the 2.3~$\mu$m and 3.6~$\mu$m methane features in Figure \ref{fig:forward_spec} and highlighted in Figure \ref{fig:ch4-contrib} are muted in the full network scenario (blue plot) compared to its reduced counterpart (orange plot) as this molecule is strongly photolysed in the upper atmosphere, with differences of the order of 200 ppm. As these differences are an order of magnitude greater than our simulated noise of JWST \citep{gardner_jwst} and the $\approx$ 20 ppm noise floors of Twinkle \citep{twinkle} and Ariel \citep{tinetti_ariel,tinetti_ariel2}, it is may be possible to infer, at least the presence of, methane photodissociation processes in the spectral data from these observatories.

\section{Retrievals}

We now evaluate the performance and biases introduced (1) when using chemical kinetics rather than equilibrium and (2) when using two different chemical kinetics networks. The simulated spectra to be fit against in the retrievals make use of the same methodology as the preceding section but are convolved with the JWST instrument response for one transit of HD\,189733\,b with NIRISS GR700XD and one with NIRSpec G395M. The error bars are obtained using the ExoWebb instrument simulator \citep{Edwards_if_he_writes_it}, which is based upon the radiometric model from \citet{terminus}. Normally HD\,189733 would saturate the NIRISS instrument, which necessitates moving the star to 27 parsecs to prevent non-linearity in the detector response. 
We utilize the same priors for all cases described in Table \ref{tab:chem-priors}. The $K_{zz}$ parameter, in particular, is fitted in the reduced and full networks cases to uniform priors between $10^3$--$10^{12}$ cm$^2$.s$^{-1}$ inclusive in log-space. For benchmark purposes and to provide comparisons with our previous works \citep{Al-Refaie_2021, Al-refaie_2021_t3.1}, we highlight below the details of our hardware setup and computing use for this work. The retrievals performed in this work do not exploit GPU acceleration, which was introduced in \citet{Al-refaie_2021_t3.1}, as the chemical kinetic solver is the dominant computational bottleneck. However, this allows us to mitigate the long computation time by exploiting large CPU-only nodes with significantly higher core counts. We use the DIRAC facility dedicating 180 cores per run for our retrieval case. The retrievals utilised the MultiNest optimizer \citep{Feroz_2009, Buchner_2016}, with 750 live points and an evidence tolerance of 0.5, resulting in around 40,000 samples. MPI was utilised to parallelise the forward model sampling of MultiNest, effectively giving a 180x sampling throughput assuming the kinetic solve takes the same amount of time for each sample. 

\begin{table}

\centering
\begin{tabular}{lll}
\hline\hline
Parameter & Prior & Range \\

\hline
$R_p$ & Uniform & 0.8--2.0 $R_{J}$ \\
$T$ & Uniform & 700.0--2500 K \\
$Z$ & log-Uniform & 10$^{-1}$--$10^{3}$ $Z_{\odot}$ \\
C/O & Uniform & 0.1--2.0\\
$K_{zz}$ & log-Uniform & 10$^{3}$--10$^{13}$ cm$^2$/s \\
\hline
\end{tabular}
\caption{Retrieval priors and ranges. The \textit{log} prefix describes fitting the ranges in log-space}
\label{tab:chem-priors}
\end{table}

\subsection{Reduced chemistry}

\begin{center}
\begin{figure}[]
    \includegraphics[width=1.0\columnwidth]{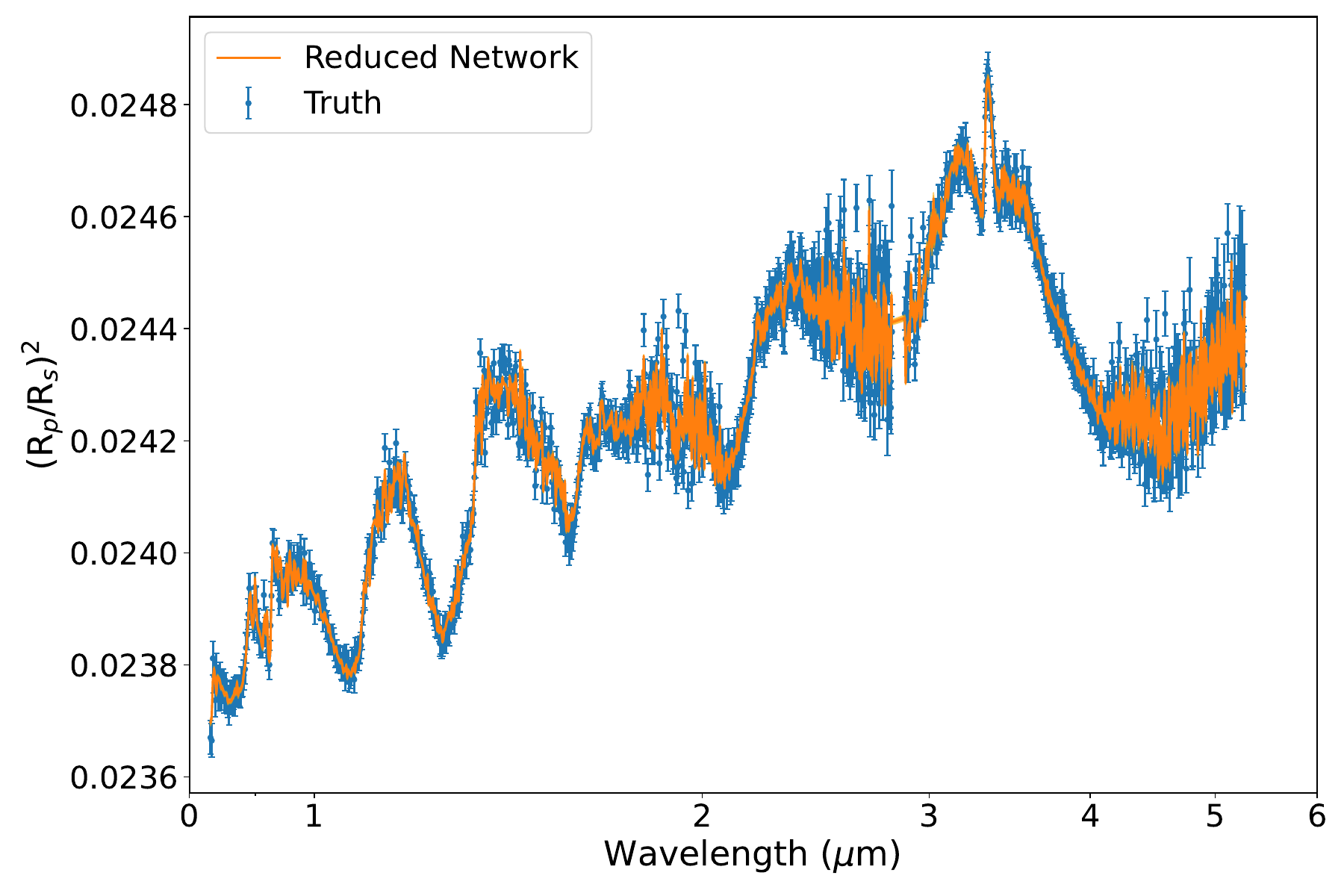}
\caption{Simulated JWST observations of HD\,189733\,b (blue) without photodissociation with retrieval best-fit models (orange) for the reduced network.}
\label{fig:reduce-nophoto-spec}
\end{figure}
\end{center}

In order to scrutinize the potential limitations and efficiencies of reduced chemical kinetic networks, we compared similarly to \cite{venot2020} but used a retrieval framework to characterize the applicability of a reduced chemical network. This test involved using a full chemical kinetic network with photodissociation disabled to simulate synthetic spectra. We then employed a reduced chemical kinetic network for the retrieval process.
Our goal was to evaluate whether the simplified network could accurately capture information of the key reactions and constituents in the atmosphere, even with the inherent simplifications it encompasses. 
Figure \ref{fig:reduce-nophoto-spec} depicts the corresponding best-fit spectra. At a glance, the best-fit spectra match well with the simulated, which corroborates the similar results in \cite{venot2020}. As illustrated in Figure \ref{fig:reduce-nophoto-post}, the posteriors are well-defined and largely lie on or near the truth, with nominal values differing by $6$--$10$\%. It's expected that the posterior mass does not align exactly with the truth as, fundamentally, the models are not exactly the same. Observing the molecular profiles in Figure \ref{fig:diseq-prof-nophoto}, many of the constituents match exactly with the truth. Exceptions are noted in the cases of H$_2$CO and HCN, which possess slight variations around the mid and upper atmospheric pressures, respectively, likely from missing reactions. HCN is of note as it's one of the targeted molecules in \cite{venot2020}. The abundances near the top of the atmosphere are higher in the full network, and that is likely due to the additional 29 HCN-producing reactions with many of their highest reaction rates in these layers for which photodissociation would act as a sink. Finally, for C$_2$H$_4$, its profile is retrieved with high precision. This is due to its involvement in the production of H and H$_2$ as either a reactant or in byproducts. In particular, its reaction pathways of C$_2$H$_5 \rightarrow$ C$_2$H$_4$ $+$ H, C$_2$H$_6 \rightarrow$ C$_2$H$_4$ $+$ H$_2$, C$_2$H$_4 \rightarrow$ C$_2$H$_3$ $+$ H and C$_2$H$_4 \rightarrow$ C$_2$H$_3$ $+$ H$_2$ have reactions rates in the order of 10$^5$~cm$^{-3}$s$^{-1}$ and tightly constrain its profile along the atmosphere.
Overall, consistent with previous studies, the reduced network possesses similar chemical information on the composition of the atmosphere against the full chemical network without photodissociation.

\subsection{Viability of full chemical kinetic retrievals.}

Due to their long solve times, full chemical kinetic networks have generally not been used in retrievals. As \textsc{FRECKLL} significantly speeds up convergence, we will assess the viability of using chemical kinetic networks in retrievals. We will assess computational viability and the associated biases using such models in retrievals. In a realistic scenario, we would not possess full information about the atmosphere and its complex processes. However, we can replicate this by using the full network with photodissociation as a proxy for our complex atmosphere and utilize the full network in retrievals to represent perfect knowledge of the system and the reduced and equilibrium chemistry as a means to quantify how our assumptions influence the retrievals. This is the same methodology as \cite{Al-refaie_2021_t3.1}

This is the first time a full disequilibrium kinetic retrieval, including vertical mixing and photochemistry, is attempted. Due to the large number of samples evaluated in atmospheric retrievals, numerical stability across the full range of parameters explored is key, highlighting the importance of the improvements described in the Methodology section. %The observed spectrum is shown in Figure \ref{fig:full-ret-spec}. 

\begin{center}
\begin{figure}[]
    \includegraphics[width=1.0\columnwidth]{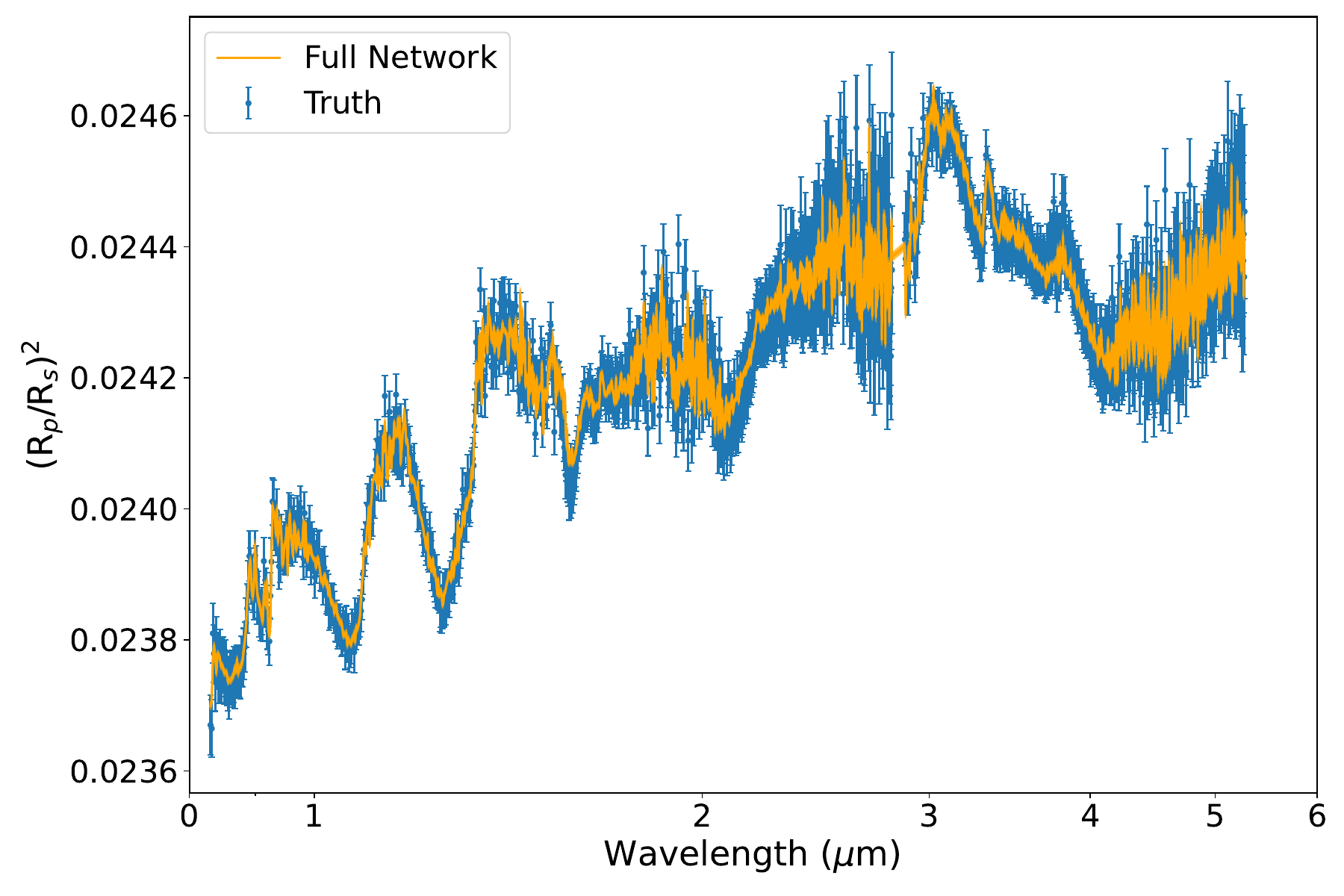}
    \includegraphics[width=1.0\columnwidth]{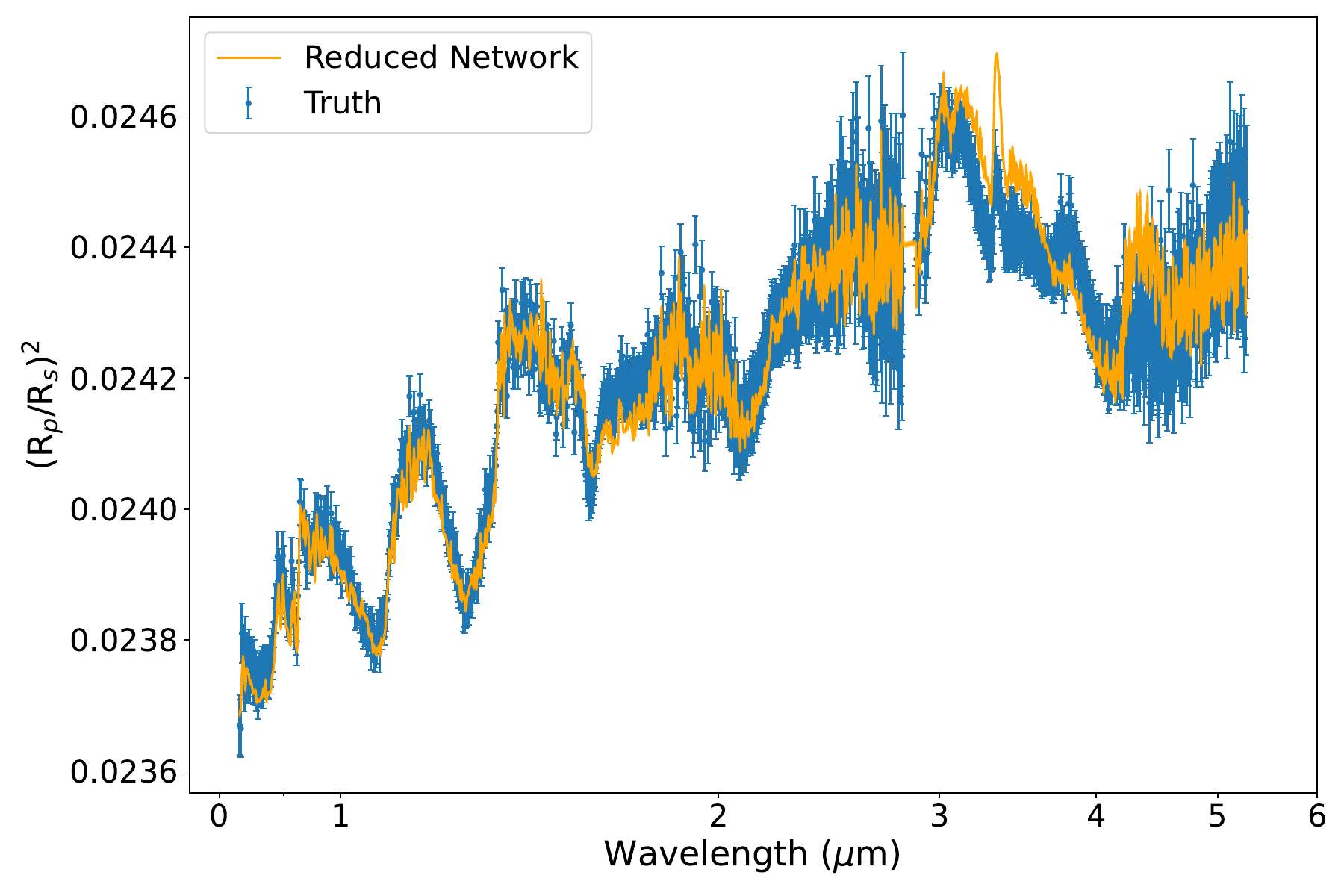}
    \includegraphics[width=1.0\columnwidth]{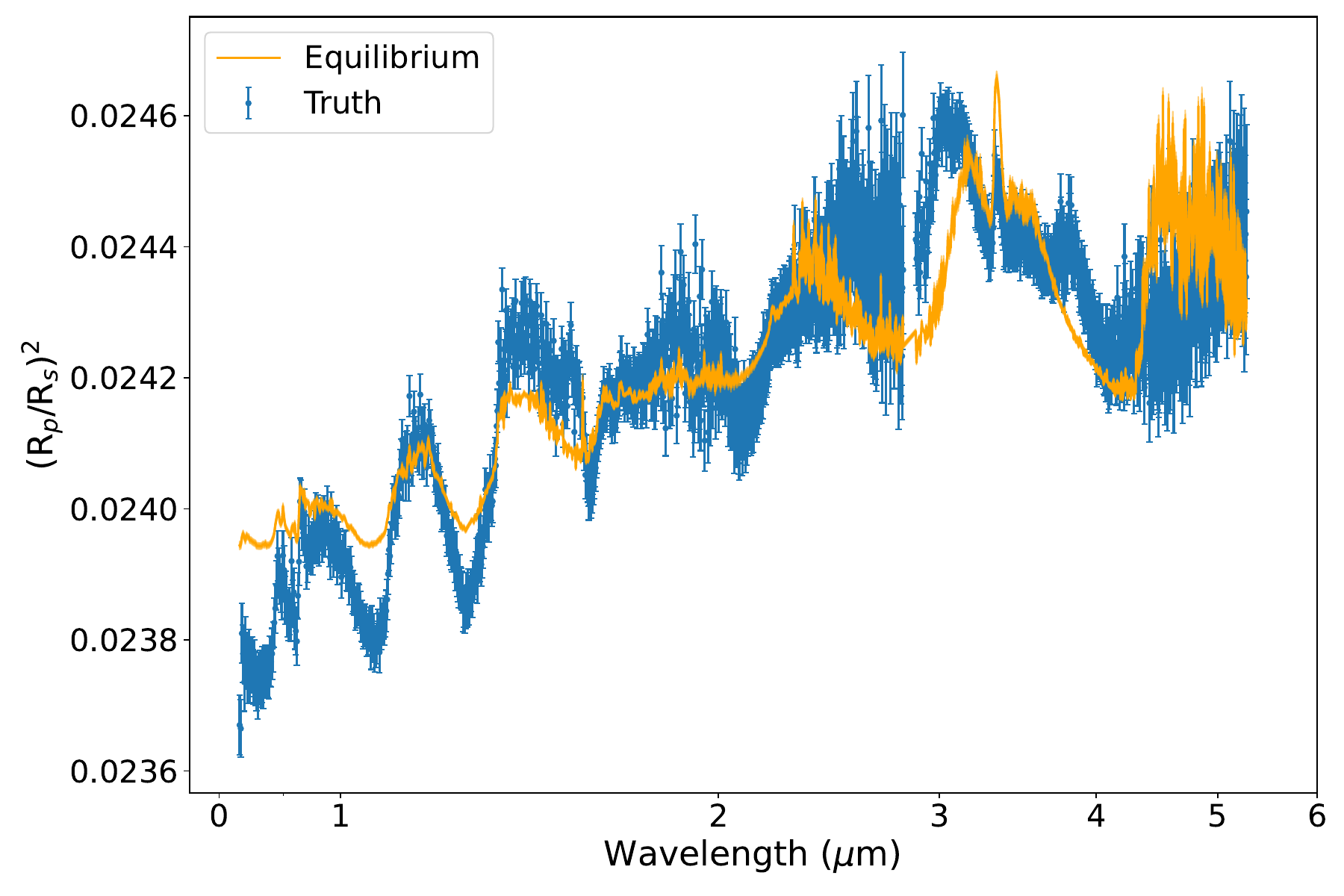}
\caption{Simulated JWST observations of HD\,189733\,b (blue) with retrieval best-fit models (orange) for the full network (top panel), reduced network (middle panel) and equilibrium (bottom panel).}
\label{fig:full-ret-spec}
\end{figure}
\end{center}

The The observed spectrum as well as the results of these retrievals are shown in Figure \ref{fig:full-ret-spec}. The kinetic runs took 8 and 24 hours to complete for the reduced and full networks, respectively. The equilibrium chemistry run took around 20 minutes. Posteriors are provided in Figure \ref{fig:diseq-post}, and the chemistry profiles are provided in Figure \ref{fig:diseq-prof}. From the inspection of the best-fit spectrum, we observe that, as expected, the full network matches the observations. Additionally, the recovered free parameters are close to the chosen true value (see Figure \ref{fig:diseq-post}), and the recovered chemical profiles match the inputs within the uncertainties (see Figure \ref{fig:diseq-prof}). 
Notably, the $K_{zz}$ has a well-defined posterior and is retrieved well by the full network on the simulated JWST observations, implying that disequilibrium processes are observable with JWST. This was also shown in previous works \citep[e.g.][]{Greene_2016, Blumenthal_2018, Molaverdikhani_2019, Drummond_2020, Venot_2020_W43}. The shape of the posterior displays a skew towards lower values, with a distinct boundary observed around 10$^{8}$ cm$^2$.s$^{-1}$. This trend can be attributed to the fact that lower $K_{zz}$ values tend to favour chemical reactions, which generally yield equilibrium-like species profiles. Importantly, $K_{zz}$ predominantly affects the mid to upper atmosphere, where it takes precedence over reaction rates and whose change in chemical composition isn't as easily probed from the JWST spectra. This assertion is supported by the extensive range of the posterior, spanning over an order of magnitude and is further corroborated by the molecular profiles shown in Figure \ref{fig:diseq-prof}. In these profiles, the uncertainties in the upper atmosphere are notably larger compared to those in the lower atmosphere, which are generally less responsive to variations in $K_{zz}$. Compared to the retrieval in the previous subsection, photochemical processes (generally not dependent on temperature) are more directly coupled to $K_{zz}$ as their reaction rates depend on the number density of species transported into this region. 
The temperature profile also displays bi-modality, which suggests a degree of degeneracy, even with perfect knowledge of the atmospheric processes. This bimodality is small with a difference between peaks of about 40~K around the truth value but suggests that higher resolution and/or lower errors on spectra are needed to fully remove degeneracies.

For the reduced network, the best-fit spectrum is at most wavelengths able to reproduce the observations, but we observe large discrepancies at certain bands. For instance, the 3.6~$\mu$m methane band is not well fitted by the reduced network, which, in our case, predicts too much methane. With an absence of photolysis reactions, the model lacks enough flexibility to compensate and deduce the correct abundance of CH$_4$ without affecting the other important molecules of the atmosphere. This is confirmed in Figure \ref{fig:diseq-prof}. In most cases, the true value in this atmosphere is outside the 1$\sigma$ predictions of the reduced retrievals. The metallicity, for instance, is found to be about Z = 6 when the input metallicity was solar (Z = 1). This is problematic as this could lead to incorrect interpretations, especially as such a parameter is commonly used to link atmospheric composition to planetary formation \citep{Oberg_2011, Moses_2013, Madhu_2016, Line_2021}.

We find similar results regarding the chemical equilibrium run. The retrieved parameters are most of the time outside the true values by more than 1$\sigma$. To compare the recovered metallicity again, assuming equilibrium chemistry, it is found to be about Z = 32. 
%In general, we find that using a chemical equilibrium assumption or a reduced scheme to recover the information content of a more complete full kinetic model that include photo-chemistry lead to strong biases. 
In general, we find that attempting to recover information content by modelling and simplifying complex processes in an atmosphere introduces strong biases in relatively unconstrained retrievals.
The recovered chemical profiles, as seen in Figure \ref{fig:diseq-prof} present large departures from the input, with the main molecules being often different by more than two orders of magnitude. The predictions for both equilibrium and reduced runs are overconfident and do not reflect the raw information content in the spectrum. This is due to the assumptions (equilibrium chemistry or pre-selected list of reactions) introduced in those models that do not capture the essential physics in the atmosphere \citep{Changeat_2019, Changeat_2020_alfnoor, Al-refaie_2021_t3.1}, in particular photodissociation. Similarly, one-dimensional retrievals, in general, may exhibit these types of biases as well \citep{Feng_2016, Caldas_2019, Taylor_2020, Changeat_2020_pc1, Macdonald_2020, Skaff_2020, Pluriel_2020}. While one might argue for the exclusive use of full chemical networks in retrievals as the most realistic, these models also face limitations. They depend heavily on the UV fluxes from the parent star and makes similar assumptions on what species are (and are not) present in the atmosphere. They are also highly dependent on our knowledge of chemical kinetics. Although considerable advances have been made recently (e.g. \citealt{Veillet2024}), certain reactions, and even certain couplings between elements (C-S or S-P for instance) are still insufficiently constrained. The approximations or assumptions made in chemical networks can, therefore, induce strong biases.  In contrast, it has been shown that simple models such as constant profiles are better at broadly retrieving the chemical composition parameters such as $Z$ and C/O \citep{Al-refaie_2021_t3.1} as they more directly fit the spectral shape of species in the retrieval. A similar approach for $K_{zz}$ could be employed by using n-layer parameterized molecular profiles \citep{Changeat_2019}, which could extract information on how species are distributed along the atmosphere. Emission spectra may also be employed in conjunction with transmission spectra to better constrain the temperature profile. Prospective work should look to understand better what information is needed to constrain particular classes of chemical models, especially when involving photodissociation. This is imperative in the era of JWST as it has already observed photolysis processes in exoplanets \citep{Tsai2023}. In summary, while complex models can theoretically provide a detailed understanding of atmospheric processes, their effectiveness is significantly diminished without sufficient data, especially in the context of wide and unconstrained retrievals. Adding information into the retrieval, such as constraining parameters or introducing more spectral data, will enhance a chemical model's ability to extract detailed information and reduce biases and degeneracies.

%Examining the retrieved molecular profiles in Figure \ref{fig:diseq-prof}, we see excellent agreement in the mid- to lower portion of the atmosphere. In particular, the truth profile for CO, H$_2$O and CO$_2$ all lie within the retrieval variances. For NH$_3$, HCN and CH$_4$ the retrieved results differ in the upper atmosphere. This is to be expected as the reduced scheme does not include photodissociation reactions from the full scheme, which are a sink for these molecules in the upper atmosphere. The retrieval does not attempt to compensate for this, which indicates that the spectrum does not contain enough information to probe these regions precisely.

\section{Conclusion}

In this work, we introduce \textsc{FRECKLL}, a cutting-edge tool designed for the rapid and stable computation and retrieval of exoplanet chemical kinetics. Central to \textsc{FRECKLL}'s efficiency is its distillation algorithm, which significantly enhances convergence to a steady state and allows for solving of large complex chemical networks in minutes. By integrating \textsc{FRECKLL} with \textsc{TauREx} 3 through its plugin system, we have for the first time successfully coupled chemical kinetics with retrievals, facilitating disequilibrium retrieval using a comprehensive kinetic network with photodissociation. We have shown that the use of strong assumptions about chemical composition (equilibrium, reduced or full networks) in retrievals could considerably bias the interpretation of observations, and we caution the reader in their use in exoplanet retrievals without significant constraints. However, this work paves the way for a new type of retrieval. If used with care, it could help improve our knowledge of exoplanetary atmospheres, particularly in the era of new telescopes beginning with the JWST. We understand the importance and benefits of open-source sharing within the academic community. However, as of the current state, \textsc{FRECKLL} is not yet available for public use. Our decision to delay the open-source release is grounded in ensuring the tool is user-friendly and free from potential pitfalls that might arise from its current intricacies. We're dedicated to refining the codebase, enhancing its documentation, and addressing any existing issues to make it robust and accessible.

\section{Acknowledgments}

This work utilised the OzSTAR national facility at Swinburne University of Technology. The OzSTAR program receives funding in part from the Astronomy National Collaborative Research Infrastructure Strategy (NCRIS) allocation provided by the Australian Government. This work utilised the Cambridge Service for Data Driven Discovery (CSD3), part of which is operated by the University of Cambridge Research Computing on behalf of the STFC DiRAC HPC Facility (www.dirac.ac.uk). The DiRAC component of CSD3 was funded by BEIS capital funding via STFC capital grants ST/P002307/1 and ST/R002452/1 and STFC operations grant ST/R00689X/1. DiRAC is part of the National e-Infrastructure. 

A.A. and Q.C. acknowledge funding from the European Research Council (ERC) under the European Union’s Horizon 2020 research and innovation programme (grant agreement No 758892, ExoAI), from the Science and Technology Funding Council grants ST/S002634/1 and ST/T001836/1 and from the UK Space Agency grant ST/W00254X/1.

Q.C. is the recipient of a 2022 European Space Agency Research Fellowship grant. 

O.V. acknowledges funding from the ANR project `EXACT' (ANR-21-CE49-0008-01), from the Centre National d'\'{E}tudes Spatiales (CNES), and from the CNRS/INSU Programme National de Plan\'etologie (PNP).

B.E. is a Laureate of the Paris Region fellowship programme which is supported by the Ile-de-France Region and has received funding under the Horizon 2020 innovation framework programme and the Marie Sklodowska-Curie grant agreement no. 945298.

The authors wish to thank Alex Thompson and Sushuang Ma for brainstorming the name of this code.

\bibliography{sample63}{}
\bibliographystyle{aasjournal}



\appendix

\renewcommand\thefigure{\thesection.\arabic{figure}}    
\renewcommand\thetable{\thesection.\arabic{table}}  

\section{Photodissociation reactions}

\setcounter{figure}{0}
\setcounter{table}{0}

Table \ref{tab:photo-reactions} shows all 55 photodissociation reactions present in the full \citet{venot2020} chemical network

\begin{table}
\scriptsize
\centering
\begin{tabular}{ccccc}
\hline\hline
Pathways & &  & Cross-Sections & Quantum yields \\ \hline
C$_{2}$H + $hv$ &$\rightarrow$ & C + C + H & \cite{fahr2003} & \cite{fahr2003} \\

C$_{2}$H$_{2}$ + $hv$ &$\rightarrow$ & C$_{2}$H + H &   \citet{cooper1995b, wu2001} & \citet{lauter2002absolute, kovacs2010h} \\

C$_{2}$H$_{3}$ + $hv$ &$\rightarrow$ & C$_{2}$H$_{2}$ + H & \citet{Fahr1998} & \citet{Fahr1998} \\

C$_{2}$H$_{4}$ + $hv$ &$\rightarrow$ & C$_{2}$H$_{2}$ + H$_{2}$ & \citet{cooper1995b}; & \citet{Holland1997}; \\
  &$\rightarrow$ & C$_{2}$H$_{2}$ + H + H & \citet{Orkin1997, wu2004temperature} & \citet{Chang1998} \\

C$_{2}$H$_{6}$ + $hv$ &$\rightarrow$ & C$_{2}$H$_{4}$ + H$_{2}$ & \citet{Au1993};  & \citet{Akimoto1965}; \\
  &$\rightarrow$ & C$_{2}$H$_{4}$ + H + H & \citet{Lee2001}; & \citet{Hampson1965}; \\
  &$\rightarrow$ & C$_{2}$H$_{2}$ + H$_{2}$ + H$_{2}$ & \citet{chen2004temperature}; & \citet{Lias1970}; \\
  &$\rightarrow$ & CH$_{4}$ + $^{1}$CH$_{2}$ & \citet{kameta1996photoabsorption} & \citet{Mount1978}\\
  &$\rightarrow$ & CH$_{3}$ + CH$_{3}$ & &  \\

C$_{2}$N$_{2}$ + $hv$ &$\rightarrow$ & CN + CN & B\'enilan et al. (in prep) & \citet{Cody1977, Jackson1979}; \\
& & & &  \citet{Eng1996}\\

CH$_{2}$CO + $hv$ &$\rightarrow$ & $^{3}$CH$_{2}$ + CO & \cite{laufer1971lowest} & Estimated \\

CH$_{3}$ + $hv$ &$\rightarrow$ & $^{1}$CH$_{2}$ + H & \citet{khamaganov2007photolysis} & \citet{Parkes1973} \\

CH$_{3}$CHO + $hv$ &$\rightarrow$ & CH$_{4}$ + CO & \cite{limao2003electronic}; & \cite{sander2006chemical}\\
  &$\rightarrow$ & CH$_{3}$ + HCO & \cite{sander2006chemical} & \\

CH$_{3}$OH + $hv$ &$\rightarrow$ & H$_{2}$CO + H$_{2}$ & \cite{burton1992absolute}; & Estimated\\
  &$\rightarrow$ & CH$_{3}$ + OH & \cite{cheng2002experimental} & \\

CH$_{3}$OOH + $hv$ &$\rightarrow$ & CH$_{3}$O + OH & \cite{vaghjiani89, matthews2005} & Estimated \\

CH$_{4}$ + $hv$ &$\rightarrow$ & CH$_{3}$ + H & \citet{Au1993}; & \citet{gans2011photolysis} \\
  &$\rightarrow$ & $^{1}$CH$_{2}$ + H$_{2}$ & \citet{Lee2001}; & \\
  &$\rightarrow$ & $^{1}$CH$_{2}$ + H + H & \citet{kameta2002photoabsorption}; & \\
  &$\rightarrow$ & $^{3}$CH$_{2}$ + H + H & \citet{chen2004temperature} & \\
  &$\rightarrow$ & CH + H$_{2}$ + H &  & \\

CHCO + $hv$ &$\rightarrow$ & CH + CO & Estimated from \cite{laufer1971lowest} & Estimated \\

CO + $hv$ &$\rightarrow$ & C + O($^{3}$P) & \citet{olney1997} & \citet{huebner1992solar} \\

CO$_2$ + $hv$ &$\rightarrow$ & CO + O($^1$D) &  \citet{huestis2011critical}; & \citet{huebner1992solar} \\
  &$\rightarrow$ & CO + O($^{3}$P) & \citet{stark2007photoabsorption}; \citet{Itya2008} & \\
  
H$_{2}$ + $hv$ &$\rightarrow$ & H + H & \citet{samson1994total}; \citet{chan1992}& Estimated \\
& & & \citet{olney1997} & \\

H$_{2}$CN + $hv$ &$\rightarrow$ & HCN + H & \cite{nizamov2003, teslja2006} & Estimated \\

H$_{2}$CO + $hv$ &$\rightarrow$ & H$_{2}$ + CO & \citet{cooper1996}; & \cite{huebner1992solar} \\
  &$\rightarrow$ & H + HCO & \citet{meller2000temperature} & \\
  
H$_2$O + $hv$ &$\rightarrow$ & H$_2$ + O($^1$D) & \citet{fillion2004}; & \citet{huebner1992solar} \\
  &$\rightarrow$ & H + H + O($^{3}$P) & \citet{mota2005}; & \\
  &$\rightarrow$ & H + OH & \citet{chan1993b}  &  \\
  
H$_{2}$O$_{2}$ + $hv$ &$\rightarrow$ & OH + OH & \cite{sander2011} & \cite{sander2011}\\

HCN + $hv$ &$\rightarrow$ & CN + H & \citet{Lee1980}; B\'enilan et al. (in prep) & \citet{Lee1980} \\

HCO + $hv$ &$\rightarrow$ & H + CO & \citet{hochanadel1980ultraviolet}; \citet{Loison1991}  & Estimated \\

HNC + $hv$ &$\rightarrow$ & CN + H & Estimated from \citet{Lee1980}; B\'enilan et al. (in prep) & Estimated from \citet{Lee1980}\\

HNO$_{2}$ + $hv$ &$\rightarrow$ & NO + OH & \cite {sander2011} & Estimated \\

HNO$_{3}$ + $hv$ &$\rightarrow$ & NO$_{2}$ + OH & \citet{sander2011} & Estimated \\

N$_2$ + $hv$ &$\rightarrow$ & N($^2$D) + N($^4$S) & \citet{samson1964, huffman1969} & Estimated\\
& & &\citet{Stark1992, chan1993c} & \\

N$_{2}$H$_{4}$ + $hv$ &$\rightarrow$ & N$_{2}$H$_{3}$ + H & \citet{Vagh1993} & \citet{Vagh1993,vaghjiani1995laser} \\

N$_2$O + $hv$ &$\rightarrow$ & N$_2$ + O($^1$D) & \cite{au1997absolute}; & \cite{okabe1978photochemistry}\\
  &$\rightarrow$ & N$_2$ + O($^1$D) & \cite{hubrich1980ultraviolet, burkholder2020chemical} &  \\

N$_{2}$O$_{3}$ + $hv$ &$\rightarrow$ & NO$_{2}$ + NO & \citet{stockwell1978near} & \citet{sander2011} \\

N$_{2}$O$_{4}$ + $hv$ &$\rightarrow$ & NO$_{2}$ + NO$_{2}$ & \citet{Vandele1998, merienne1997} & \citet{sander2011} \\

NH$_{2}$ + $hv$ &$\rightarrow$ & NH + H & \cite{HUEBNER2015} & \cite{HUEBNER2015}\\

NH$_{3}$ + $hv$ &$\rightarrow$ & NH$_{2}$ + H & \citet{Burton1993, chen1998}; & \citet{McNesby1962} \\
& & & \citet{cheng2006} & \\

NO + $hv$ &$\rightarrow$ & N$_{4}$S + O($^{3}$P) & \citet{Iida1986, chan1993a} & \citet{huebner1992solar} \\

NO$_{2}$ + $hv$ &$\rightarrow$ & NO + O($^{3}$P) & \cite{au1997absolute}; & \cite{huebner1992solar}\\
  &$\rightarrow$ & NO + O($^1$D) & \cite{vandaele2002high} & \\

NO$_{3}$ + $hv$ &$\rightarrow$ & NO$_{2}$ + O($^{3}$P) & \cite{sander1986temperature, yokelson1994temperature}; & \cite{HUEBNER2015}\\
  &$\rightarrow$ & NO + O$_{2}$ & \cite{orphal2003ultraviolet}& \\
  
OH + $hv$ &$\rightarrow$ & O($^1$D) + H & \citet{huebner1992solar} & \citet{vandish1984} \\

OOH + $hv$ &$\rightarrow$ & OH + O($^{3}$P) & \cite{sander2011} & \cite{sander2011} \\
\hline
\end{tabular}
\caption{The 55 photodissociation reactions, their associated cross-sections and quantum yields included in the \citet{venot2020} chemical network.}
\label{tab:photo-reactions}
\end{table}

\clearpage 

\section{Posterior Distributions Derived from Simulations Using the Full Chemical Network (No Photodissociation)}

\setcounter{figure}{0}

\begin{center}
\begin{figure}[h]
    \includegraphics[width=0.9\linewidth]{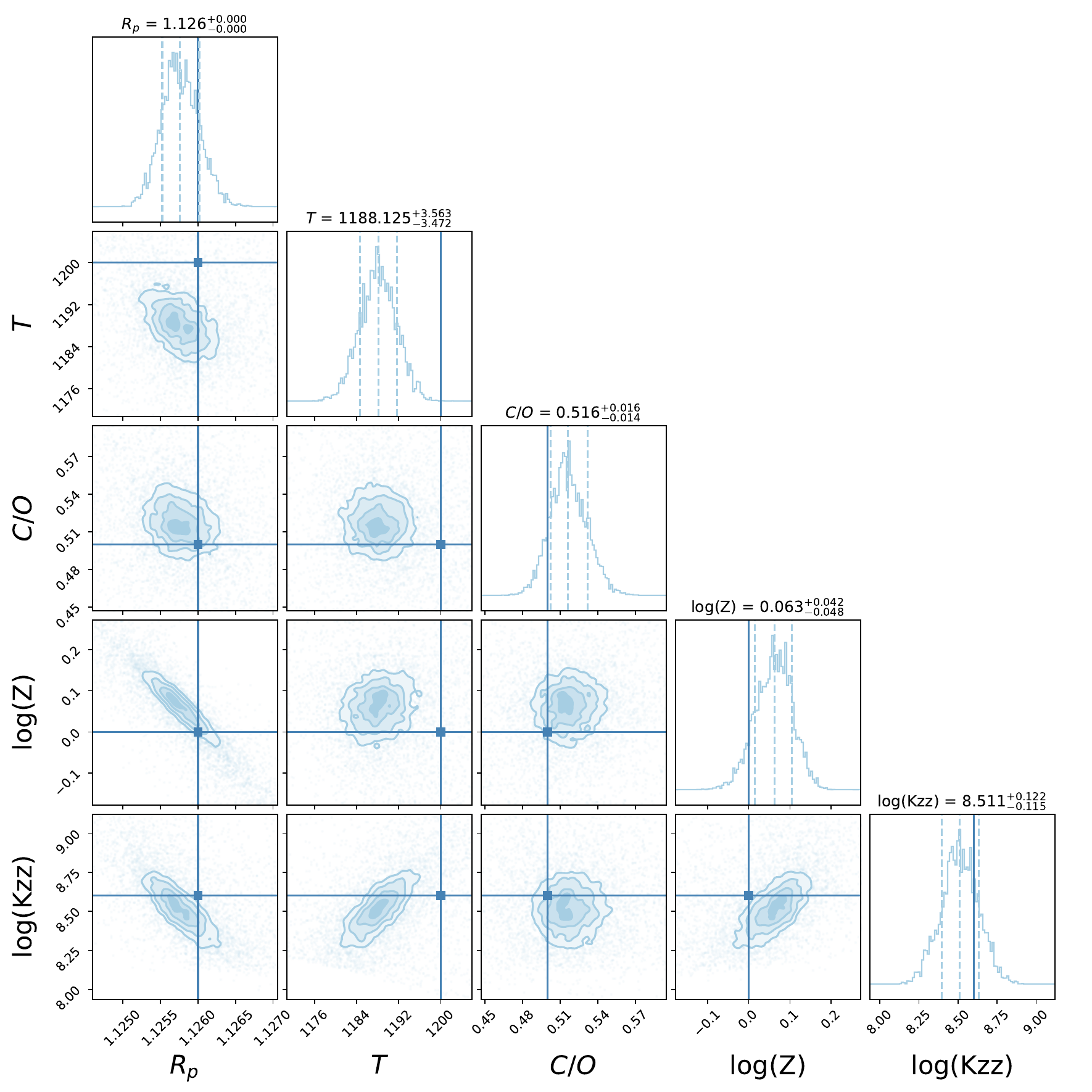}
\caption{Posterior distributions from the retrieval analysis, where the simulated JWST spectra of HD\,189733\,b—generated using the full chemical network from \cite{venot2020} without photochemistry—are retrieved using the reduced chemical network. Parameters for the simulations can be referenced in Table \ref{tab:hd189-test}. The light blue line denotes the true values as listed in Table \ref{tab:hd189-test}. }
\label{fig:reduce-nophoto-post}
\end{figure}
\end{center}

Figure \ref{fig:reduce-nophoto-post} shows the posterior distributions for the simulated JWST spectra retrieved with the reduced chemical network.

\clearpage

\section{Molecular profiles obtained in retrievals (no photodissociation)}

\setcounter{figure}{0}   

\begin{center}
\begin{figure}[h]
    \includegraphics[width=0.9\linewidth]{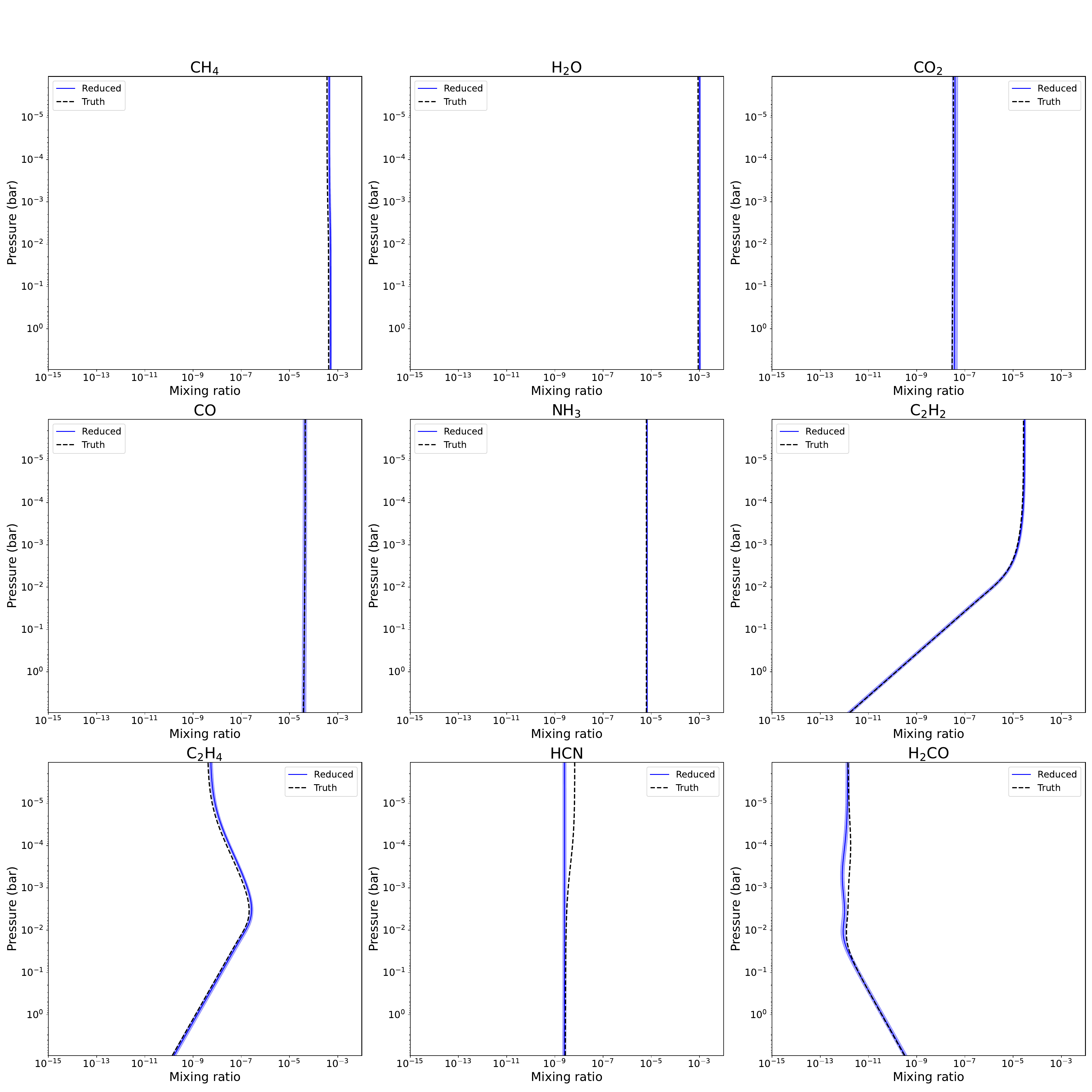}
\caption{Chemical abundances profiles recovered by the reduced chemical network retrievals in our simulations of HD\,189733\,b. Shaded regions are 1$\sigma$ confidence intervals.}
\label{fig:diseq-prof-nophoto}
\end{figure}
\end{center}

Figure \ref{fig:diseq-prof-nophoto} shows the abundance profiles of the main chemical species in the reduced chemical network retrievals.

\clearpage

\section{Posterior Distributions Derived from JWST Spectrum Simulations Using the Full Chemical Network, Including Photodissociation}

\setcounter{figure}{0}   

\begin{center}
\begin{figure}[h]
    \includegraphics[width=0.9\linewidth]{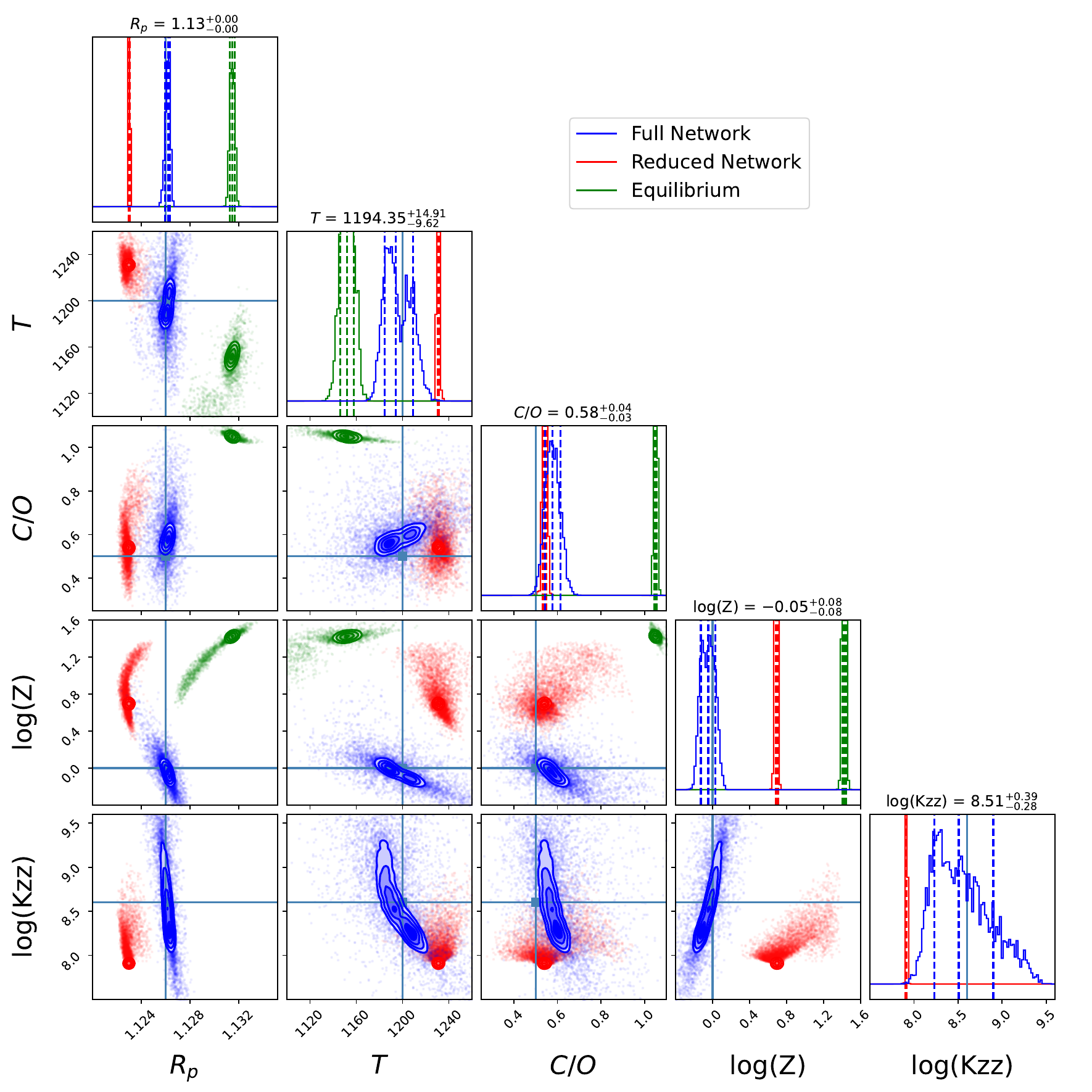}
\caption{Posterior distributions derived from the retrieval analysis of simulated JWST spectra for HD\,189733\,b, where the spectra were generated using the full chemical network from \cite{venot2020}. Retrievals were performed using the full chemical network, the reduced network, and the equilibrium chemistry as described by \citep{Agundez2012}. Parameters guiding these simulations and retrievals can be found in Table \ref{tab:hd189-test}. In the displayed posteriors, the blue and red curves represent the results from retrievals using the full and reduced networks with \textsc{FRECKLL}, respectively, while the green curve showcases results from the equilibrium chemistry-based retrieval. The light blue line indicates the true values as referenced in Table \ref{tab:hd189-test}. Values on the top of the posterior are from the full network retrieval.}
\label{fig:diseq-post}
\end{figure}
\end{center}

Figure \ref{fig:diseq-post} shows the posterior distributions for the simulated JWST retrievals.

\clearpage

\section{Molecular profiles obtained in retrievals (photodissociation)}

\setcounter{figure}{0}   

\begin{center}
\begin{figure}[h]
\begin{interactive}{animation}{full_II_retrieval.mp4}
    \includegraphics[width=1.0\linewidth]{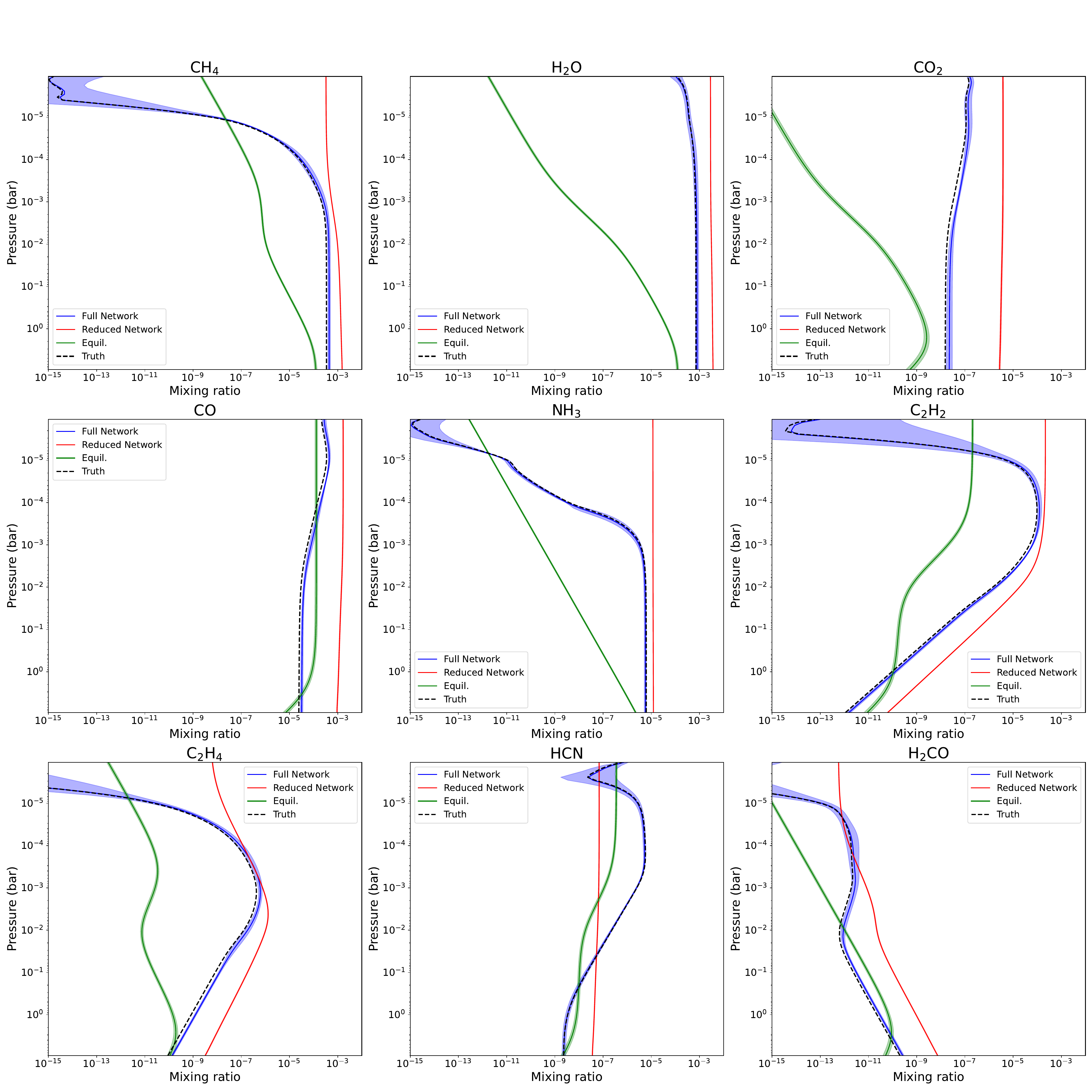}
\end{interactive}
\caption{Chemical abundances profiles recovered by the reduced (red), full (blue) and equilibrium (green) retrievals in our simulations of HD\,189733\,b. Shaded regions are 1$\sigma$ confidence intervals. The animated version of this plot shows the time evolution of the best-fit full chemical network retrieval from $t=0$ until steady-state.}
\label{fig:diseq-prof}
\end{figure}
\end{center}

Figure \ref{fig:diseq-prof} shows the abundance profiles of the main chemical species in the retrievals.

\end{document}